\documentclass[a4paper]{article}

\usepackage[utf8]{inputenc}
\usepackage[T1]{fontenc}

\usepackage[a4paper]{geometry}
\usepackage{amsmath,stackengine,mathtools,amssymb}
\usepackage{amsthm} 
\usepackage{comment}
\usepackage{lipsum}
\usepackage[usenames,dvipsnames]{xcolor}
\usepackage{minitoc}
\usepackage{tikz}
\usepackage{filecontents}
\usepackage[numbers,sort&compress]{natbib}
\usepackage[colorlinks=true]{hyperref} 
\hypersetup{urlcolor=blue, citecolor=red} 
\usepackage{tikz} 
\usetikzlibrary{matrix} 
\usepackage{graphicx}
\usepackage[english]{babel}
\usepackage{hyperref}
\usepackage{amsfonts}
\usepackage{empheq} 
\usepackage{tabularx}
\usepackage{appendix}
\usepackage{color} 
\usepackage{lineno} 
\usepackage{float} 
\usepackage{caption} 
\usepackage{xcolor}   
\usepackage{environ}  
\usepackage{enumitem}       
\usepackage{booktabs}       
\usepackage{diagbox} 
\usepackage{array} 
\usepackage{natbib} 

\newtheorem{definition}{Definition}[section]
\newtheorem{oss}{Remark}[section]
\newtheorem{example}{Example}[section]

\newcommand{\numberset}{\mathbb}
\newcommand{\N}{\numberset{N}}
\newcommand{\R}{\numberset{R}}
\newcommand{\curlyQ}{\mathcal{Q}}

\newcommand{\curlyY}{\mathcal{Y}}
\DeclareMathOperator*{\argmin}{argmin}

\newcommand{\h}[1]{\widehat{#1}}

\newcommand{\pt}{\partial_t}
\newcommand{\px}{\partial_x}

\usepackage{bigfoot} 
\usepackage[numbered,framed]{matlab-prettifier} 

\usepackage{filecontents} 

\lstMakeShortInline"

\usepackage{siunitx} 
\DeclareSIUnit[]\mass
{\text{\ensuremath{M}}}

\makeatletter
\let\@fnsymbol\@arabic
\makeatother
\title{{A model for membrane degradation using a gelatin invadopodia assay}} 
\author{Giorgia Ciavolella
	\thanks{INRIA Bordeaux-Sud-Ouest, Institut de Mathématiques de Bordeaux, CNRS UMR 5251 \& Université de Bordeaux, 351 cours de la Libération, 33405 Talence Cedex, France}
	\thanks{Corresponding author: giorgia.ciavolella@inria.fr, ORCID number 0000 0002 6614 3966}
	\and
	Nathalie Ferrand\thanks{Sorbonne Universit\'{e} Cancer Biology and Therapeutics, INSERM, CNRS, Institut Universitaire de Canc\'{e}rologie, Saint- Antoine Research Center (CRSA), F-75012, Paris, France}
	\and
	Michèle Sabbah\footnotemark[3] 
	\and
	Beno\^ \i t Perthame\thanks{Sorbonne Universit\'{e}, Inria, Universit\'{e} de Paris, Laboratoire Jacques-Louis Lions, UMR7598, 4 place Jussieu, 75005 Paris, France}
	\and
	Roberto Natalini \thanks{Istituto per le Applicazioni del Calcolo 'M.Picone', Rome, Italy}
} 

\date{\today}

\begin{document} 
	
	\maketitle
	
	\begin{abstract} 
		One of the most crucial and lethal characteristics of solid tumors is represented by the increased ability of cancer cells to migrate and invade other organs during the so-called metastatic spread. This is allowed thanks to the production of matrix metalloproteinases (MMPs), enzymes capable of degrading  a type of collagen abundant in the basal membrane separating the epithelial tissue from the connective one. {In this work, we employ a synergistic experimental and mathematical modelling approach to explore the invasion process of tumor cells. 
			A mathematical model composed of reaction-diffusion equations describing the evolution of the tumor cells density on a gelatin substrate, MMPs enzymes concentration and the degradation of the gelatin is proposed. This is completed with a calibration strategy. We perform a sensitivity analysis and explore a parameter estimation technique both on {synthetic} and experimental data in order to find the optimal parameters that describe the \textit{in vitro} experiments. A comparison between numerical and experimental solutions ends the work.}		
	\end{abstract}
	
	\begin{flushleft}
		\noindent{\makebox[1in]\hrulefill}
	\end{flushleft}
	\begin{flushright}
		\noindent{\makebox[1in]\hrulefill}
	\end{flushright}

	\section{Introduction}
	
	Tumors are complex diseases characterised by high diversity and incidence, Sung \textit{et al.}~\cite{sung}. One of the most crucial and lethal processes is represented by the increased ability of cancer cells to migrate and invade other organs during the so-called \textit{metastatic spread}, Dillek\aa s \textit{et al.}~\cite{dillekas}. In the primary tumor, cancer cells could acquire useful mutations that allow them to penetrate different barriers and to disseminate themselves into secondary organs. Then, we observe the transition from an \textit{in situ} stage to an invasive one. It is now well known that metastatic cancer cells typically move in clusters which have greater predisposition of forming metastasis than single cells, Aceto \textit{et al.}~\cite{aceto}, Bubba \textit{et al.}~\cite{bubba}, Hong \textit{et al.}~\cite{hong}.
	
	Invasion of cells through layers of extracellular matrix is a key step in tumor metastasis, inflammation, and development. The process of invasion involves several stages, including adhesion to the matrix, degradation of proximal matrix molecules, extension and traction of the cell on the newly revealed matrix, and movement of the cell body through the resulting gap in the matrix, Friedl and Wolf~\cite{friedl}. Each of these stages of invasion is executed by a suite of proteins, including proteases, integrins, GTPases, kinases, and cytoskeleton-interacting proteins.
	One of the most difficult barriers for cells to cross is the basement membrane, also composed by {extracellular matrix (ECM)}. This kind of membrane separates the epithelial tissue from the connective one 
	and, among its functions, we can distinguish a supportive role and an isolating one.
	Tumor cell invasion is associated with an enhanced capability of tumor cells to degrade ECM. Tumor cells produce several hydrolytic enzymes including matrix metalloproteinases (MMPs).
	This process is, for instance, observed for breast tumors, as considered in our study.
	
	Many questions concerning invasion details remain unanswered. Over the last decade, the research interest on this process is increasing in order to highlight the main cues with the aim of controlling and treating the phenomenon (Chaplain \textit{et al.}~\cite{giverso}, Ciavolella \textit{et al.}~\cite{ciadapou}, Franssen \textit{et al.}~\cite{franssen}, Gallinato \textit{et al.}~\cite{gallinato},  Giverso \textit{et al.}~\cite{giverso2}). 
	
	In particular, we are interested in better describing one of the main biological phenomenon responsible for the invasion one, which is membrane degradation. In fact, \textit{in vitro} invasion investigations (using the XCELLigence technology, Connolly and Maxwell \cite{connolly}, Ke \textit{et al.}~\cite{ke}, Martinez-Serra \textit{et al.}~\cite{martinez}, Obr \textit{et al.}~\cite{obr}, Turker \textit{ et al.}~\cite{turker}, Zaoui \textit{ et al.}~\cite{zaoui}) are not able to give details on {this process}. Consequently, we build here a mathematical model which describes degradation of an ECM-like biological membrane through the production of MMPs by cancer cells. At the same time, we provide numerical simulations with a sensitivity analysis followed by a parameters estimation study. The work is completed by experimental results consisting of cells seeded in wells containing at their bottom fluorescent gelatin. We focus the attention on breast cancer, the most common malignancy among women. {The  Triple Negative Breast Cancer (TNBC) patients are mainly treated with combinations of  chemotherapy with severe side effects and afrequent recurrence of the metastasis.} If non-metastatic, it has high {chances of healing, but, on the contrary,} advanced breast cancer with metastasis are considered incurable with the {present} therapies, Harbeck \textit{et al.}~\cite{harbeck}, Waks and Winer~\cite{waks}.
	
	The outline of the paper is as follows. 
	In Sections~\ref{sec:mathmodel} and~\ref{sec:mathmodelnondim}, we introduce the mathematical model and its dimensionless form.  Section~\ref{sec:parameter} is divided in two subsections describing the sensitivity analysis and the parameter estimation problem. 	{In Subsection~\ref{subsec:sensitivity}, we briefly present  the concept of {sensitivity} analysis and then the results on our model. }The same is done in Subsection~\ref{subsec:inversepb}, in the case of both {synthetic} data and experimental ones. The subsection ends with numerical simulations which are compared with the biological experiments. Finally, in Section~\ref{sec:conclusions}, we conclude the paper presenting also some perspectives that both improve our work and develop it. At the end of the paper, the reader can find the Supplementary Materials~\ref{app:discretisation} in which we give more details concerning the numerical method behind the simulations presented.

	
	\section{Mathematical model}\label{sec:mathmodel}
	{We build a mathematical model describing the invasive behaviour of tumor cells on a gelatin coated plate. Their movement is influenced by the gelatin, since their primary role is to degrade it, due to MMPs enzymes. The major hypothesis is that cells degrade gelatin below them and, after, they move to degrade around. Experiments capturing this behaviour are realised using the QCM™ Gelatin Invadopodia Assay.
		Tumor cells are plated on a culture surface coated with a thin layer of green fluorescent gelatin and they are filmed over several days. Videos can show both cells movement and the consequent creation of holes in the gelatin.
	}
	
	In accordance with this kind of experimental setting, we consider a domain $\Omega$ representing a top view of a single plate.  {We do not include the third dimension because the thickness of the gelatin layer is not significant compared to cells size. Consequently, the movement is only realised on a surface. } For simplicity, we take the rectangular two-dimensional {domain $\Omega$}.
	For $x\in \Omega$, $t>0$, we consider a three species system for  cells density $u(x,t)$, MMPs enzymes concentration $m(t,x)$, and the damage function $d(t,x)\in [0,1]$, related to the amount of gelatin $q(t,x)=1-d(t,x)\in [0,1]$. Equations write as
	\begin{equation}\label{eq:syst}
		\left\{\begin{array}{ll}
			\pt u=div(D(d)\nabla u), & \mbox{ in } \Omega,\\[1ex]
			\pt m=D_m \Delta m + \beta (1-d)u-\alpha m, & \mbox{ in } \Omega,\\[1ex]
			\pt d=\gamma m (1-d), & \mbox{ in } \Omega,
		\end{array}
		\right.
	\end{equation}
	where 
	\begin{equation*}\label{eq:diff}
		D(d)=D_L d+D_G(1-d)= D_G+(D_L-D_G)d.
	\end{equation*}
	We impose no-flux boundary conditions on
	$\partial \Omega$ for $u$ and $m$, \textit{i.e.}
	\begin{equation*}\label{eq:N}
		\left\{\begin{array}{ll}
			{D(d)\nabla u\cdot \textbf{n}} =0,\\[1ex]
			D_m \nabla m \cdot \textbf{n}=0,
		\end{array}
		\right.
	\end{equation*}
	where \textbf{n} is the outward unit normal at the boundary.
	We complete the system with the following initial conditions
	\begin{equation*}\label{eq:init_data}
		\left\{\begin{array}{ll}
			u(0,x)=u_0(x),\\[1ex]
			m(0,x)=m_0(x)=0,\\[1ex]
			d(0,x)=d_0(x)=0,
		\end{array}
		\right.
	\end{equation*}
	with $u_0$ a random function on $\Omega$.
	
	\textbf{Equation for $u$.} The first equation in \eqref{eq:syst} describes the evolution of the cells density. The diffusion coefficient depends on the diffusion $D_L>0$ into the liquid and $D_G>0$ on the gelatin, {with ${D_L >D_G}$}.  {We do not consider the case ${D_L = D_G}$}, otherwise we would have a standard diffusion equation. We observe that when the gelatin is intact ($d=0$, then $q=1$), cells move randomly on the gelatin, whereas when it is completely destroyed ($d=1$, then $q=0$), cells diffuse into the liquid.
	The expression of $D(d)$ is not new in applications. In the literature we can find it, for example, in models describing the chemical transformation of calcium carbonate stones under the attach of sulphur dioxide, Aregba-Driollet \textit{et al.}~\cite{aregba}, Guarguaglini and Natalini~\cite{guarguaglini}.
	
	\textbf{Equation for $m$.} The second equation is a reaction-diffusion like equation for the MMPs with diffusion coefficient $D_m>0$, production rate $\beta>0$ and death rate $\alpha>0$. In particular, production of MMPs is due to cells and to the fact that they sense the gelatin below them. Moreover, we assume that MMPs diffuse locally and, consequently, we add the condition 
	\begin{equation*}
		\sqrt{\frac{D_m}{\alpha}} \ll 1,
	\end{equation*}
	where $\sqrt{\frac{D_m}{\alpha}}$ is {homogeneous to $\sqrt{x^2}$ and corresponds} to the diffusion length.
	
	\textbf{Equation for $d$.} Finally, the equation for the damage derives from the damage mechanics, Kachanov~\cite{kachanov}. It can be written in terms of the gelatin $q$ as 
	\[\pt q=-\gamma m q,\]
	which has an exponential decreasing solution. The damage is produced at rate $\gamma$ by the MMPs $m$.
	{
		\begin{oss}
			The choice of a macroscopic model is dictated by the high number of tumor cells in experiments. Moreover, it is more interesting to look at the general behaviour of cells instead of the single individual. Indeed, tracking them, we could observe that over ${72}$ hours they make local movements around the initial position and at very low velocities. No particular interactions between cells could then be taken into account. Finally, a hybrid approach could be introduced, but this would largely increase the number of parameters, adding complexity in  their estimation. We deduce the multiple advantages of our simple PDE model that is able to describe degradation phenomena.
	\end{oss}}
	
	\subsection{Dimensionless model}\label{sec:mathmodelnondim}
	Before analysing the model, we propose a nondimensional form. It has several advantages, as the reduction of the number of parameters and the fact that their units are unimportant, see Murray~\cite{murray1}, Segel~\cite{segel}. 
	Upon changes of time and space variables
	\begin{equation*}
		\tilde t=\alpha t, \quad \tilde x= \sqrt \frac{\alpha}{D_m} x,
	\end{equation*}
	and appropriate scaling for $u, m,$ and $d$, namely
	\begin{equation*}
		\begin{array}{l}
			u(t,x)=\tilde u \left(\alpha t, \sqrt{\frac{\alpha}{D_m}} x\right),\\[1.5ex]
			m(t,x)= \frac \alpha \gamma \tilde m \left(\alpha t, \sqrt{\frac{\alpha}{D_m}} x\right),\\[1.5ex]
			d(t,x)= \frac{D_m}{D_L-D_G} \tilde d \left(\alpha t, \sqrt{\frac{\alpha}{D_m}} x\right),
		\end{array}
	\end{equation*}
	we find a parametrised version, again with homogeneous Neumann boundary conditions. For simplicity in notation, eliminating the tilde, we find
	\begin{equation}\label{eq:nondim}
		\left\{\begin{array}{ll}
			\pt  u=div((\theta +  d)\nabla  u), & \mbox{ in } \Omega,\\[1ex]		
			\pt  m= \Delta  m + k_1 (1-p d) u-  m, & \mbox{ in } \Omega,\\[1ex]
			\pt  d=\frac 1 p  m (1-p d), & \mbox{ in } \Omega,
		\end{array}
		\right.
	\end{equation}
	where $\theta=\frac{D_G}{D_m}, p=\frac{D_m}{D_L-D_G}, k_1= \frac{\gamma\beta}{\alpha^2}$. Here, $D( d)= \theta +  d$.
	We assume all parameters positive, except $p$ that satisfies the restriction $p\leq 1$, since we require $1-pd\geq 0$.

	%

	\section{Sensitivity analysis and parameter estimation}\label{sec:parameter}
	
	{We provide  a parameters estimation study for the dimensionless Model~\eqref{eq:nondim}.  Indeed, parameters are not accessible by direct measurements from experiments.
		To test the estimation technique, it is interesting to work at first with synthetic data.  The idea is that we create a {numerical (synthetic)} solution, through our {Model~\eqref{eq:nondim}, using parameters from the literature.} Then, we find {the best set of parameters that gives reasonable solutions} and we can evaluate the error made in the choice of parameters, see Subsection~\ref{subsec:inversepb}.} 
	
	Another useful study that can be done in parallel is the sensitivity analysis, see Subsection~\ref{subsec:sensitivity}. It provides us with an instrument to examine how the choice of the parameters affects the model dynamics. This is also a key indicator in the case in which the error between one of the {optimal parameters and its corresponding initial estimate} is too big. In fact, the sensitivity analysis could show us that some parameters do not have important effects on solutions. Thus, a huge error on their estimation is not so important. 
	
	{At the end of Subsection~\ref{subsec:inversepb}, we provide an example of possible experimental data {and we reiterate the consolidated procedure performed with synthetic data}. We evaluate the optimal parameters describing the biological experiment and we provide numerical simulations of the model solutions.}
	{Except for this experiments paragraph, in the following we integrate the theory with our model analysis
		considering parameters 
		taken} from Di Costanzo~\textit{et al}~\cite{dicostanzo}, and Braun \cite{braunphd}, 
	\begin{equation*}
		\begin{array}{c}
			{D_L=\SI{7d-7}{\cm\squared\per\second}, \quad D_G= \SI{d-7}{\cm\squared\per\second}, \quad D_m= \SI{5d-7}{\cm\squared\per\second},}
		\end{array}	 
	\end{equation*} 
	{and from } Franssen \textit{et al} \cite{franssen},
	\begin{equation*}
		{\alpha=\SI{2.5d-6}{\per\second}, \quad \beta=
			\SI{4.9d-6}{\mass\per\second}.}
	\end{equation*}
	Consequently, the nondimensional parameters are
	\begin{equation}\label{eq:parameters}
		\theta= 0.2, \quad p=0.83,  \quad k_1= 0.78.
	\end{equation}
	
	\subsection{Sensitivity analysis }\label{subsec:sensitivity}
	The lack of {parameters} availability from experiments imposes an uncertainty in the output of the model. To obtain as reliable results as possible, we have to study the influence of the parameters on the model dynamics through a \textit{local sensitivity analysis}. Local methods are the simplest and the most common. They are based on a one-factor-at-a-time (OAT) method. It consists in perturbing one parameter at a time to see the effect on the output for each parameter of the model. Sensitivity analysis (SA) is then performed monitoring changes in the output  through, for example, a derivative-based approach.
	
	{SA determines dependencies of input parameters $\curlyQ=(\curlyQ_1,...,\curlyQ_n)$ and output of the model~$\curlyY$. In the following we illustrate  the procedure.
		The measure of the influence of the parameter $\curlyQ_i$ on the output $\curlyY(\curlyQ)$ is calculated by the partial derivative with respect to this parameter. Since the model parameters differ by several orders of magnitude, we introduce a normalisation with respect to the mean value. Using the finite difference approach, the sensitivity of the output {with respect to} the parameter $\curlyQ_i$ is obtained as
		\begin{equation*}
			S= \frac{\curlyY(\curlyQ_1,...,\curlyQ_i\pm\delta,...,\curlyQ_n)-\curlyY(\curlyQ_1,...,\curlyQ_n)}{\curlyY(\curlyQ_1,...,\curlyQ_n)} \;\frac{\curlyQ_i}{\delta},
		\end{equation*}
		where $\delta=0.05\, \curlyQ_i$ to consider a $5\%$ perturbation of the parameters.
	}

	\paragraph{For our model.} 
	
	We report here the results obtained for our Model~\eqref{eq:nondim}, using the {{dimesionless} parameters from the literature} in~\eqref{eq:parameters}. We have evaluated the sensitivity of parameters $\curlyQ=[\theta, p, k_1]$ {taking as output $\curlyY$ the maximum value at the final time for cells density and the total mass of enzymes and gelatin at the final time. It is not interesting to consider the cells total mass since this is a conserved quantity.} {
		The OAT method does not examine the whole range of values of the input parameters as the global SA. However, this method is easier to define and computationally more efficient than the global sensitivity, for which different techniques are possible, see Saltelli \textit{et al.}~\cite{saltelli}.  }
	
	{In Table~\ref{tab:sensitivity}, we show for each perturbed parameter $\curlyQ_i\pm\delta$, the sensitivity $S_\curlyY$ related to the output $\curlyY$.  
		Sensitivity should be less than one or around it but not much bigger in order to have low sensitivity of the model {with respect to} parameters. 
		{Since in our case it is at maximum around $1$, we deduce that solutions have not a great influence on big changes in parameters. Of course, the bigger is the sensitivity, the bigger is the influence on solutions.} Moreover, from Table~\ref{tab:sensitivity}, we deduce that  the diffusion parameter does not play a remarkable role in the behaviour of enzymes $m$ and degradation $d$, whereas it is the most 'critical' parameter for cells evolution. In this case, in fact, $p$ and $k_1$ are of order $10^{-2}$. More critical is the influence of $p$ on degradation and $k_1$ on both degradation and enzymes concentration, but still sensitivity is good enough. From SA, we infer that {changing {the literature} parameters} do not greatly affect solutions behaviour. Not knowing real solutions of the model, this is a good estimate to know about.

		\captionsetup[table]{labelfont=bf,textfont={it}} 
		\begin{table}[H]
			\centering
			\begin{tabular}{c c c c c c}
				\toprule
				\diagbox{$\curlyQ_i\pm\delta$}{$S_\curlyY$}  & $S_{\mbox{max} u}$ & $S_{\mbox{mass} m}$  & $S_{\mbox{mass} d}$  \\
				\cmidrule(r){1-1}\cmidrule(lr){2-2}\cmidrule(lr){3-3}\cmidrule(lr){4-4}
				$\theta+\delta$   
				& $0.6262$ & $\SI{6.57d-4}{}$ & $\SI{7.05d-5}{}$ \\[1ex]
				$\theta-\delta $    & $0.6705$ & $\SI{7.03d-4}{}$ & $\SI{7.48d-4}{}$\\[1ex]
				$p+\delta$ & $0.0141$ & $\SI{2.44d-4}{}$ & $0.9524$ \\[1ex]
				$p-\delta$
				& $0.0155$ & $\SI{2.7d-6}{}$ & $1.0526$ \\[1ex]
				$k_1+\delta$  & $0.0146$ & $0.9980$ & $0.9982$ \\[1ex]
				$k_1-\delta$ & $0.01469$ & $0.9982$ & $0.9972$\\[1ex]
				\hline
			\end{tabular}
			\caption{We collect sensitivity values $S_\curlyY$ {per} each parameter $\curlyQ_i\pm\delta$, with $\delta=0.05\, \curlyQ_i$ . $S_{\mbox{max} u}$ is the sensitivity with output the maximum of cell density at the final time {($72$ hours)}, $S_{\mbox{mass} m}$ is the sensitivity with output the amount of enzymes again {after $72$ hours} and the same for $S_{\mbox{mass} d}$ {with respect to} gelatin degradation. }
			\label{tab:sensitivity}
		\end{table}

		\subsection{Inverse problem: parameter estimation with {synthetic} and biological data}\label{subsec:inversepb}
		With the knowledge that changing our parameters {from the literature} we do not modify too much solutions (Subsection~\ref{subsec:sensitivity}), we can now estimate, through an inverse problem, the good parameters useful to recover {reasonable solutions compared to synthetic and} experimental {ones}.

		A convenient reformulation of the inverse problem is to write it as a minimization problem.
		\begin{definition}
			Let $F(\curlyQ)=\curlyY$ be the forward problem with $\curlyQ =(\curlyQ_1,...,\curlyQ_n)$ the parameters and $\curlyY$ the numerical solution found with the finite difference scheme. The inverse problem is defined as a minimisation problem of the form 
			\begin{equation}\label{eq:minimisation}
				\curlyQ_{opt}:= \argmin_{\curlyQ\in \R^n} \|F(\curlyQ)-\curlyY_{exp}\|_{L^2}^2,
			\end{equation}
			where $Q_{opt}$ are the estimated parameters and $\curlyY_{exp}$ the experimental {(or synthetic)} data.
		\end{definition}
		Unfortunately, inverse problems are usually ill-posed. Then, initial small perturbations can lead to large ones in the results. In other words, small errors between the solution of the forward problem and the experimental data can lead to arbitrary large errors between the given parameters and the estimated ones. Hence the necessity of a regularisation method to compute a stable approximation of the minimiser, Braun~\cite{braunphd, braunarticle}.
		{Thus, we extend the minimisation in~\eqref{eq:minimisation} with a Tikhonov regularisation as
			\begin{equation*}
				\curlyQ_{opt}:= \argmin_{\curlyQ\in \R^n} ( \; \|F(\curlyQ)-\curlyY_{exp}\|_{L^2}^2 + \lambda\|\curlyQ-\curlyQ_0\|_{L^2}^2 \; ),
			\end{equation*}
			where $\lambda>0$ is the regularisation parameter and the
			\textit{a priori} estimate $\curlyQ_0\in\R^n$ represents an \textit{a priori} knowledge about the solution.
		}
		
		{From gelatin invadopodia assay, it is possible to extract information mainly concerning tumor cells, since it could be difficult to quantify the gelatin and the holes created by degradation. One of the reasons is that cells have their own fluorescence and this could alter the quantification of gelatin. Up to our knowledge, it is very difficult to have data on enzymes.
			Thus, we take $\curlyY=u$ and $\curlyY_{exp}=u_{exp}$.
			Ignoring $\curlyQ_0$ due to lack of {\it a priori} information on parameters, we infer that
			\begin{equation}\label{eq:invpb}
				\curlyQ_{opt}:= \argmin_{\curlyQ\in \R^n} ( \; \|F(\curlyQ)-u_{exp}\|_{L^2}^2 + \lambda\|\curlyQ\|_{L^2}^2 \; ). 
			\end{equation}
			Thus, }if an unknown noise is included in the data, there might be different solutions that minimise $\|F(\curlyQ)-u_{exp}\|^2_2$, but among these solutions only the one which minimises $\|\curlyQ\|^2_2$ is selected. {In our case, the main interest will be in minimising the error among solutions. Thus, in the following we will choose very small values of $\lambda$ such that the norm of $\curlyQ$ will not play a significant role.}

		\paragraph{Parameter estimation with {synthetic} data for our model.} 
		
		Before working on biological data, it is commonly used to test the estimation on the {synthetic} data.
		Therefore, we develop the procedure used to estimate parameters for Model~\eqref{eq:nondim} taking the parameters from the literature. Choosing the latter as in~\eqref{eq:parameters}
		\[\curlyQ_{lit} \quad=\quad [\theta=0.2, \qquad p=0.83, \qquad k_1=0.78],\]
		we derive numerical solutions $u, m$ and $q=1-d$. 
		{Synthetic solutions are created from the numerical solutions of the model with the literature parameters. We choose two different synthetic data: at first, we use as $u_{exp}$ the exact numerical solution, and in a second step we consider a perturbation of it. Indeed, in both cases, the minimisation of the functional in~\eqref{eq:invpb} starts with $F(\curlyQ)$, the numerical solution obtained with the perturbed literature parameters, and thus we aim to recover the best parameters such that $F(\curlyQ)$ is close to $u_{exp}$. So at first with $u_{exp}$ the numerical solution of our Model~\eqref{eq:nondim}, it should be easier to recover them, while adding a small perturbation, the idea is to reproduce an error that we also encounter with experimental data. } 
		
		Thus, we perturb parameters with { $n =0.1$, $0.2$ or $0.4$} Gaussian noise, namely the perturbed parameters $\curlyQ_{pert}$ are composed by
		\begin{equation*}\label{eq:parampert}
			{\theta_{pert}=\theta+\omega_\theta \, n \; \theta,\qquad
				p_{pert}=p+\omega_p \, n \; p,\qquad k_{1_{pert}}=k_1+\omega_{k_1} \, n \; k_1,}
		\end{equation*}
		with $\omega_\theta$ and $\omega_{k_1}$ random numbers in $[-1,1]$, whereas $\omega_p$ is again a random number in $[-1,1]$ in the case {$n=0.1, 0.2$ and $[-1,0]$ if $n=0.4$}. Indeed, we recall that $p\leq 1$, so that $1-pd\geq 0$. 
		Then, we look for the best parameters that solve the inverse Problem~\eqref{eq:invpb} in an interval  $I_{lit} = [\curlyQ_{lit}-0.5 \curlyQ_{lit}, \curlyQ_{lit}+0.5 \curlyQ_{lit}]$ and starting with the perturbed parameters above.
		So, we evaluate the parameters that minimise the error $E(\curlyQ)$ between our {synthetic} data and the new solution $\h{u}$ obtained with one of the parameters $\curlyQ$ in the research interval $I_{lit}$ plus the Tikhonov regularisation with $\lambda=10^{-6}$, namely
		\begin{equation*}\label{eq:errorestimate}
			E(\curlyQ)= \|\h{u}(t)-u_{exp}(t)\|_2^2 + \lambda \|\curlyQ\|_2^2, \quad \mbox{ for } t=48 \mbox{ hours}.
		\end{equation*}
		{ The choice of the time $t=48$ hours is dictated by the dissipative behaviour of diffusion problems. Indeed, we can either loose information on parameters looking at the final time or not having them at all considering the initial time (initial data do not depend on parameters).
		}
		Finally, we evaluate the relative error between the estimated parameters $\curlyQ_{opt}$ and the initial ones, \textit{i.e.}
		\begin{equation*}\label{eq:errparam}
			E_\theta=\frac{\|\theta-\theta_{opt}\|_2}{\|\theta\|_2}, \qquad E_p=\frac{\|p-p_{opt}\|_2}{\|p\|_2}, \qquad E_{k_1}=\frac{\|k_1-k_{1_{opt}}\|_2}{\|k_1\|_2}.
		\end{equation*}
		We also estimate the relative error
		between the  {synthetic} data $u_{exp}$ and the solution with the estimated parameters, that is defined as
		\begin{equation}\label{eq:errorsolutionparameterestimated}
			E_{\curlyQ_{opt}}(t)=\frac{\|u_{exp}(t)-u_{\curlyQ_{opt}}(t)\|_2}{\|u_{exp}(t)\|_2}, \quad \mbox{ for } t=0,24,48,72 \mbox{ hours}.
		\end{equation}
		
		
		\noindent\textit{{Synthetic} data: numerical solution}
		
		\vspace{3mm}
		
		At first, we choose as {synthetic} data the numerical solution $u$.  {This means that the numerical solution of our model, found with the parameters from the literature, simulates the experimental data available. Thus, to find the optimal parameters with~\eqref{eq:invpb}, we substitute $u_{exp}$ with the numerical solution obtained with parameters from the literature and $F(\curlyQ)$ with the numerical solution obtained with the set of optimal parameters found at each iteration{, starting with the perturbed parameters}.}

		\begin{example}\label{ex4:paramestimated}
			We consider a $10\%$ perturbation on parameters $\curlyQ_{lit}=[\theta, p, k_1]=[0.2, 0.83, 0.78]$, randomly extracting $\omega_\theta=-0.48$, $\omega_p= -0.49$ and $\omega_{k_1}=0.51$. We deduce 
			\[\curlyQ_{pert}=[0.19 ,\quad 0.871 ,\quad 0.82].\]
			Solving the inverse problem in~\eqref{eq:invpb}, we infer that
			\[\curlyQ_{opt}= [0.199,\quad 0.415,\quad 0.39].\]
			{The errors on parameters are ${E_\theta=10^{-6}, \, E_p=0.5,\, E_{k_1}=0.5}$, whereas \[E_{\curlyQ_{opt}} < 8 \times 10^{-6}.\]}
			
			Unfortunately, errors on $p$ and $k_1$ are high.
			However, as expected from the sensitivity analysis in Subsection~\ref{subsec:sensitivity}, this does not impact solutions.
			In Figure~\ref{fig:errorsex3} left, we show the error $E_{\curlyQ_{opt}}$ between the data $u$ and the solution with estimated parameters. {The same is provided also for the gelatin $q$. Both are below $ \SI{8d-6}{}$ indicating a good approximation of solutions. This confirms that a property of our model is stability.}
			
		\end{example}
		
		\begin{oss}\label{oss:ex56}
			{The same results are obtained considering higher perturbations on parameters.}
		\end{oss}
		
		We conclude the example with a final observation. The relative error between solutions is of the {same order as} $\lambda=10^{-6}$, and also lower. This implies that the minimisation does not consider anymore the norm of solutions, but only the norm of parameters $\curlyQ$, see Equation~\eqref{eq:invpb}. This is why we end up with a big error on parameters: the algorithm is trying to find very small parameters. Then, considering a lower $\lambda=10^{-12}$, we could force the minimiser to decrease only {the norms of the solutions}. The results are {in the following}. We take the same data as before and, solving the inverse problem, we obtain 
		\[\curlyQ_{opt}=[0.2, \quad 0.809, \quad 0.761].\]
		The errors on parameters are significantly smaller, \textit{i.e.}
		\[{E_\theta= 10^{-8} , \quad E_p=0.02, \quad E_{k_1}=0.02}.\]
		Again, we have a good estimate of solutions as before, see Figure~\ref{fig:errorsex3} right.
		
		\captionsetup[figure]{labelfont=bf,textfont={it}}
		\begin{figure}[H]
			\centering
			\includegraphics[scale=0.35]{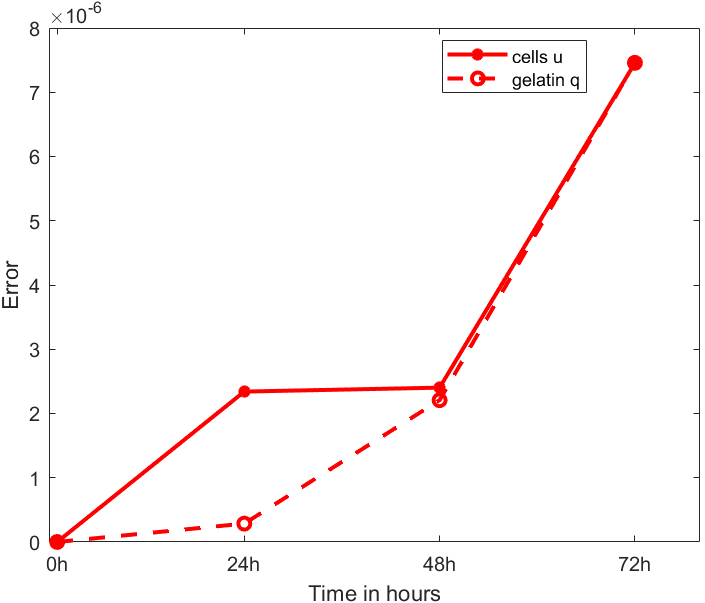} \quad
			\includegraphics[scale=0.35]{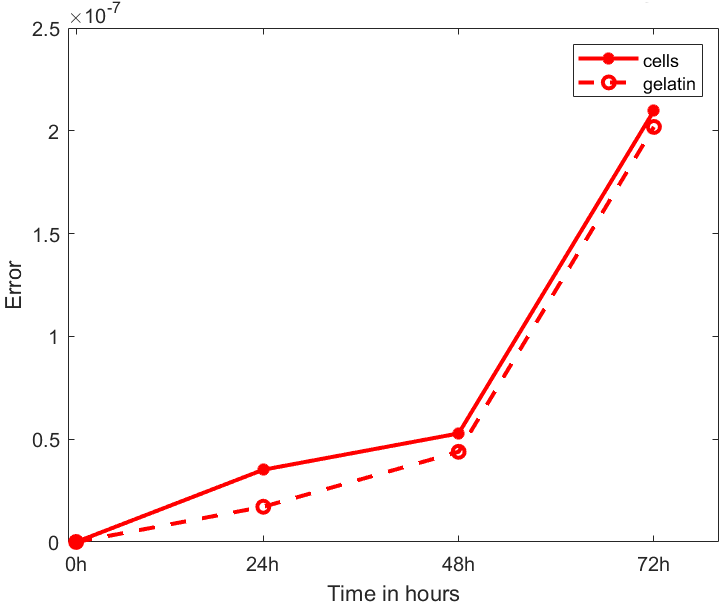}
			
			\vspace{-2mm}
			
			\caption{{On the $x$-axis, the time in days until $3$ days, which represents the duration of the experiments. On the $y$-axis, the error $E_{\curlyQ_{opt}}$ in~\eqref{eq:errorsolutionparameterestimated}, both for cells $u$ and gelatin $q$  (dash line). The graphs correspond to the case in which the synthetic data are the numerical solutions, but the regularisation parameter $\lambda$ changes. Indeed, on the left, $\lambda= 10^{-6}$, whereas on the right $\lambda= 10^{-12}$.}}
			\label{fig:errorsex3}
		\end{figure}

		\vspace{3mm}
		
		\noindent\textit{{Synthetic} data: perturbed numerical solution}
		
		\vspace{3mm}

		Experimental  (non {synthetic}) data have always some unknown noise due for instance to instrumentation. This is why in the following we consider a more realistic setting in which the {synthetic} data are a perturbation of the numerical solution of Model~\eqref{eq:nondim}.
		We perturb with $5\%$ Gaussian noise the numerical solution, \textit{i.e. } {$u_{pert}=u+\omega_u \, 0.05 \; u,$ where $\omega_u$ is a random matrix with entries in $[-1,1]$. This Gaussian noise is with zero mean, thus perturbed solutions are  on average the exact numerical ones.}
		
		\begin{example}\label{ex1:paramestimated}
			We consider a $10\%$ perturbation on parameters $\curlyQ_{lit}=[\theta, p, k_1]=[0.2, 0.83, 0.78]$. Randomly extracting $\omega_\theta= -0.11$, $\omega_p= -0.15$ and $\omega_{k_1}= -0.83$, we deduce
			\[\curlyQ_{pert}=[0.198, \quad 0.817, \quad 0.715].\]
			Solving the inverse problem in~\eqref{eq:invpb}, we get
			\begin{equation*}
				\curlyQ_{opt}=[0.199, \quad 0.415, \quad 0.539].
			\end{equation*}
			{This infers an error on parameters such that ${E_{\theta}= 0.0007 ,\, E_{p}=0.5,\, E_{k_1}= 0.3}$, whereas \[E_{\curlyQ_{opt}}< 3\times 10^{-3}.\]}
			
			As before, errors on $p$ and $k_1$ are high. 
			In Figure~\ref{fig:errorsex4} left, we show the errors $E_{\curlyQ_{opt}}$, which are for both $u$ and $q$ below the initial $5\%$ error  between the {synthetic data and the numerical solution.}
			
			\begin{oss}
				We do not get very satisfactory results in terms of parameter estimation. {However, the most important characteristic is that solutions obtained with the estimated parameters $\curlyQ_{opt}$ do not greatly differ from the experimental data (the perturbed solutions in this {synthetic} case). In contrast, we remark that the functional to be minimised in~\eqref{eq:invpb} does not include {either $q$ or $m$}.  Indeed, having more information also on gelatin and enzymes concentration could allow to include them in the functional, thus bringing us good estimations on the other parameters. 
				}
				In particular, the parameter $p$ is present both in the equation for enzymes concentration $m$ and gelatin degradation $d$, then we should need both of them to obtain a good estimate on $p$. The parameter $k_1$ is only in the equation for the enzymes and, indeed, we can approximate it having experimental data on enzymes. 
			\end{oss}
			
			\begin{oss}
				A different situation promises to be the case with {higher perturbation on parameters. Indeed}, we notice a better convergence rate for $k_1$ {with respect to} Example~\ref{ex1:paramestimated}, probably due to a better formulation of the minimisation problem which well represents the final minima. Otherwise, we always have good final errors on solutions.
			\end{oss}
			
		\end{example}
		
		So if we look for the optimal parameters describing the {synthetic} experimental data (chosen as the perturbed numerical solution $u_{pert}$), we deduce:
		\begin{itemize}
			\item a good approximation of $\theta$ with an error around {$0.001 $};
			\item a maximal error on $p$; 
			\item a varying error on $k_1$, which is around {$0.04$ or $0.3$ }in the examples presented.
		\end{itemize}
		{Moreover, we have good error values $E_{\curlyQ_{opt}}$} between the experimental synthetic solution and the one found with the optimal parameters. Indeed, for both cells density and gelatin concentration we obtain $E_{\curlyQ_{opt}} \sim 3\%$. 
		
		In order to obtain accurate optimal parameters, we should need either more experimental data, to be included in the functional definition, or eventually a model reduction. Concerning $E(\curlyQ)$, we could modify the value of $\lambda$, maybe a lower one, {as in} Example~\ref{ex4:paramestimated} where the error between solutions is of the {same order as} $\lambda$. 
		Regarding a model reduction, even if relatively simple, System~\eqref{eq:nondim} could still be too complex to describe experiments. A detailed analysis of it could highlight particular behaviours that we currently ignore. 
		We could also add an {\it a priori} knowledge on parameters such as $\curlyQ_0=\curlyQ_{lit}$. This last addition would be less realistic but, at least, we force the minimisation to consider parameters not too far away from the {\it a priori}  ones. An example is shown just below.
		\begin{example}
			Let $\lambda=10^{-3}, \curlyQ_0=\curlyQ_{lit}$.
			We consider a $10\%$ perturbation on parameters $\curlyQ_{lit}=[\theta, p, k_1]=[0.2, 0.83, 0.78]$, randomly extracting $\omega_\theta= 0.27$, $\omega_p= -0.12$ and $\omega_{k_1}= -0.81$.
			Solving the inverse problem in~\eqref{eq:invpb}, we get
			\begin{equation*}
				\curlyQ_{opt}=[0.2, \quad 0.806, \quad 0.803].
			\end{equation*}
			{This infers an error on parameters ${E_{\theta}= 0.003, \, E_{p}=0.03, \, E_{k_1}= 0.03,}$ whereas \[E_{\curlyQ_{opt}} < 3\times 10^{-3}.\]}
			In Figure~\ref{fig:errorsex4} right, we show the errors $E_{\curlyQ_{opt}}$, which are for both $u$ and $q$ below the initial error of $5\%$.

		\end{example}
		\captionsetup[figure]{labelfont=bf,textfont={it}}
		\begin{figure}[H]
			\centering
			\includegraphics[scale=0.35]{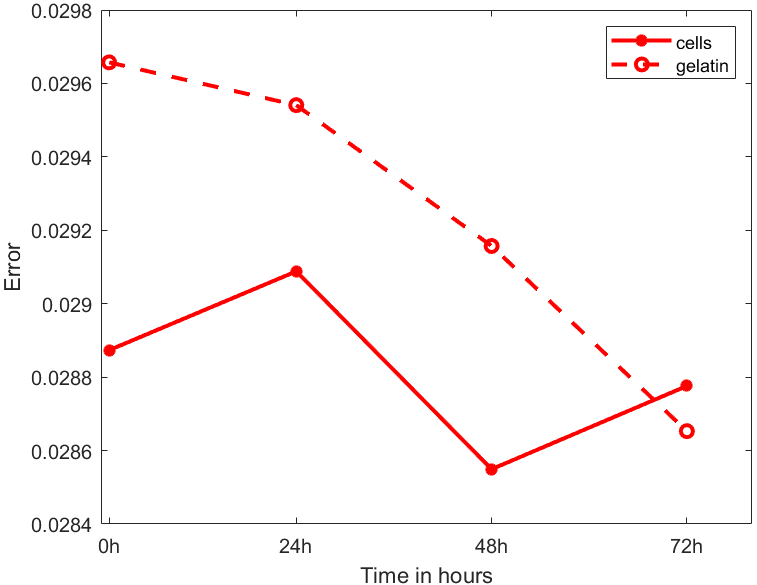}\quad
			\includegraphics[scale=0.35]{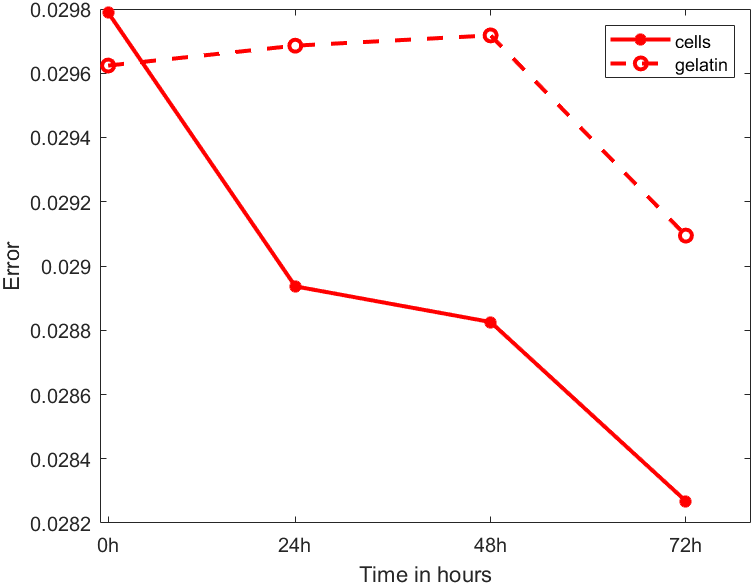}
			
			\vspace{-2mm}
			
			\caption{{On the $x$-axis, the time in days until $3$ days, which represents the duration of the experiments. On the $y$-axis, the error $E_{\curlyQ_{opt}}$ in~\eqref{eq:errorsolutionparameterestimated}, both for cells $u$ and gelatin $q$  (dash line). The graphs correspond to the case in which the synthetic data are the perturbed numerical solution, but the a priori information of parameters changes. On the left, we lack in information, whereas
					on the right $\curlyQ_0=\curlyQ_{lit}$. There are no important variations in the solutions error. Instead a difference can be seen in the estimated parameters.}}
			\label{fig:errorsex4}
		\end{figure}

		\paragraph{Parameter estimation with biological data for our model.} 
		{ 
			We briefly present the experimental setting.
			The study is on the breast carcinoma cell line called MDA-MB-231 {with a triple negative breast cancer (TNBC) phenotype}. As introduced before, cells are placed on  a thin layer of green fluorescent gelatin. Gelatin means to mimic basal membrane which has a thickness of $10$ to $\SI{300}{\nano\meter} $ which is smaller that the size of a cell, \textit{i.e.} $\SI{10}{\micro\meter} $.
			Fluorescence is useful to distinguish areas in which gelatin has been consumed. Such assays have also revealed that invasive cells extend small localised protrusions, called \textit{invadopodia}, {where MMP enzymes are localised and} {from which membrane degradation starts}. Pioneered by Wen-Thien Chen in the 1980's (\cite{chen3,chen2,chen1}),
			visualisation of invadopodia ECM degradation by fluorescent gelatin has emerged as the most prevalent technique for evaluating cellular invasive potential, Artym \textit{et al.}~\cite{artym}, Martin \textit{et al.}~\cite{martin}.  }
		
		{Experimental data consist of videos over $72$ hours on a portion of the plate, measuring 
			{2530$\times$2530}\,\si{\micro\meter\squared}.
			They focus either on cells movement or on gelatin degradation. {An example of the results is given in Figure~\ref{fig:73}. At the beginning, gelatin is intact (top image right) and {every} day the consumption increases and this corresponds to the {darker regions}. On the left column, the configuration of tumor cells starting from the initial one on the top till the third day (on the bottom).}
			
		}
		
		In possession of the biological data, we can apply the parameter estimation tested before {considering $u_{exp}$ as} the {biological }experimental data, instead of {the synthetic ones}. 
		{We have decomposed the experimental video in $18$ images (one {every }four hours). Using the Matlab tool \textit{impixel}, it is possible to manually identify cells position and to have a discrete representation of them. Thus,} we need to transform particles into densities in order to obtain a macroscopic view of cells, that we name $u_{exp}$. This procedure consists of centring a Gaussian kernel on each cell, {see Braun~\cite{braunphd, braunarticle}}.  Then, discretising $u_{exp}$, we can compare it with our numerical solution through the inverse problem written in~\eqref{eq:invpb}{, that having reliable data only on cells, can be rewritten as}
		\begin{equation*}\label{eq:invpb2}
			\theta_{opt}:= \argmin_{\theta} ( \; \|F(\theta)(t)-u_{exp}(t)\|_{L^2}^2 + \lambda\|\theta\|_{L^2}^2 \; ),
		\end{equation*}
		{with $t$ corresponding to the values at which we have evaluated $u_{exp}$, \textit{i.e.} {every }$4$ hours in the interval $[0,72]$ hours.
			We highlight that we are not looking for optimal parameters $p$ and $k_1$, since they do not significantly influence the behaviour of cells. Indeed, sensitivity analysis with experimental data shows very similar results to the ones in Table~\ref{tab:sensitivity}{, thus $p$ and $k_1$ play very little role on cells dynamics }.
			Consequently, they are going to be fixed.
			Furthermore, we realised that parameters from the literature were not appropriate to describe experimental data. 
			The first problem is in the description of cells movement that is clear only with a higher $\theta$. Moreover, $k_1$ is too low to allow enzymes production and the consequent gelatin degradation.
			Thus, we modify them as in the following. We take $p$ and $k_1$  fixed to $p=3\times 10^{-2}$ and $k_1=780$, whereas we set as  $[10,150]$ the minimisation interval for $\theta$. We test several initial random values for $\theta \in [30,55]$. Setting the regularisation parameter $\lambda=10^{-9}$, in Figure~\ref{fig:thetaoptexp} we show the minimisation results.}
		
		\begin{figure}[H]
			\centering
			\includegraphics[width=.55\textwidth]{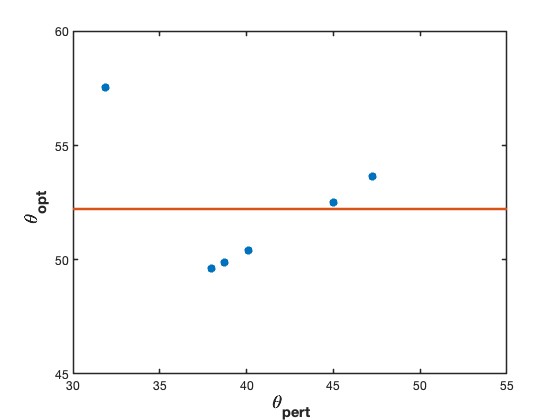}	
			\vspace{-2mm}
			\caption{{The optimal parameter $\theta$ found varying the initial perturbed one in the interval $[30,55]$. We have selected $6$ representative values and we can observe that the optimal $\theta$ is well confined in a small interval. Taking the mean value, we have that $\theta_{opt}=52.23$.}}
			\label{fig:thetaoptexp}
		\end{figure}
		{The relative error in~\eqref{eq:errorsolutionparameterestimated} between the experimental solution and the numerical one with estimated parameter $\theta$ {for} fixed $p$ and $k_1$ is always around $0.4${, which is quite good when dealing with biological data}. In this case, as for the functional, we consider Expression~\eqref{eq:errorsolutionparameterestimated} for the experimental times considered, \textit{i.e.}  {every }$4$ hours in the interval $[0,72]$ hours.}
		
		{
			In Figure~\ref{fig:solnumcell}, we show  an example  of {the numerical solution for the cells density (on the left) compared with the experimental one represented through the particle to density transformation (on the right)}. We have considered the numerical solution $u$ for Model~\eqref{eq:nondim} with  the mean optimal parameter $\theta=52.23$, and with $p=3\times 10^{-2}$ and $k_1=780$. In Figure~\ref{fig:solnumgel72}, we also propose a comparison between the numerical solution of the gelatin degradation (on the left) and the experimental data (on the right) at the final time. In this case, the fitting is much different, since apparently not all cells have produced the enzymes that degrade the gelatin and it was not possible to quantify gelatin consumption from experiments. }
		\section{Conclusions and perspectives}\label{sec:conclusions}
		
		{In this work, we set up a {preliminary} mathematical model to describe gelatin degradation as a result of the action of MMPs enzymes produced by tumor cells. This process is strictly related to cancer cells invasion through a thin membrane. The model presented describes the evolution in time of tumor cells, MMPs enzymes and gelatin degradation through a macroscopic view. }
		
		{The model is combined with a sensitivity analysis and a parameter estimation. 
			Sensitivity analysis highlights how perturbation on parameters does not affect solutions of the PDE system. Indeed, it confirms the {stability of our model}. Then, we have performed parameter estimation both on synthetic data and on biological ones. Despite the lack of a complete biological knowledge, we were able to build an appropriate description of degradation experiments. As it can be observed in Figure~\ref{fig:solnumcell}, a very attractive perspective is to test our model looking at cells in the entire experimental domain and on different set of data. Indeed, we could both better capture the diffusive behaviour given by our model and have more
			information 
			on enzymes activity and degradation, the two main poorer information that influenced the research of the complete optimal set of parameters. A deeper study on how enzymes are produced by tumor cells could be helpful, since in the experimental data proposed only a small amount of holes has appeared, see Figure~\ref{fig:solnumgel72}.  } 
		
		{More complex models considering the action of other cells involved in the invasion process, such as fibroblasts~\cite{malaquin} could be very interesting to analyse. Moreover, as stressed during the paper, the gelatin is a very thin layer. A two-dimensional approach is then required. However, this work could be a first step in studying not only basal membrane degradation, but also ECM one. ECM is in fact a thicker layer in which cells move thanks to degradation.
			Furthermore, the understanding of the degradation process is crucial in modelling invasion. Indeed, the estimated parameters helps to quantify membrane permeability inside the Kedem-Katchalsky membrane conditions.
			Finally, a more mathematical analysis on Model~\eqref{eq:nondim} can reveal interesting behaviour of solutions.}

		
		\section*{Acknowlegments}
		G.C. has received fundings from the European Research Council (ERC) under the European Union’s Horizon 2020 research and innovation programme (grant agreement No $740623$). The work of G.C. was also partially supported by GNAMPA-INdAM.
		The authors are grateful to Elishan Christian Braun for fruitful discussions, concerning the numerical implementation of the parameter estimation method. Finally, authors are very grateful to reviewers for their appropriate and constructive remarks.
		
		\section*{Statements and Declarations}
		{{\bf Competing Interests} The authors declare that they have no competing interests.}
		
		\section*{Data availability}
		All data supporting the findings of this study are available within the paper and its Supplementary Information.
		
		\captionsetup[figure]{labelfont=bf,textfont={it}}
		\begin{figure}[H] 	
			\begin{center}
				\includegraphics[width=.75\textwidth]{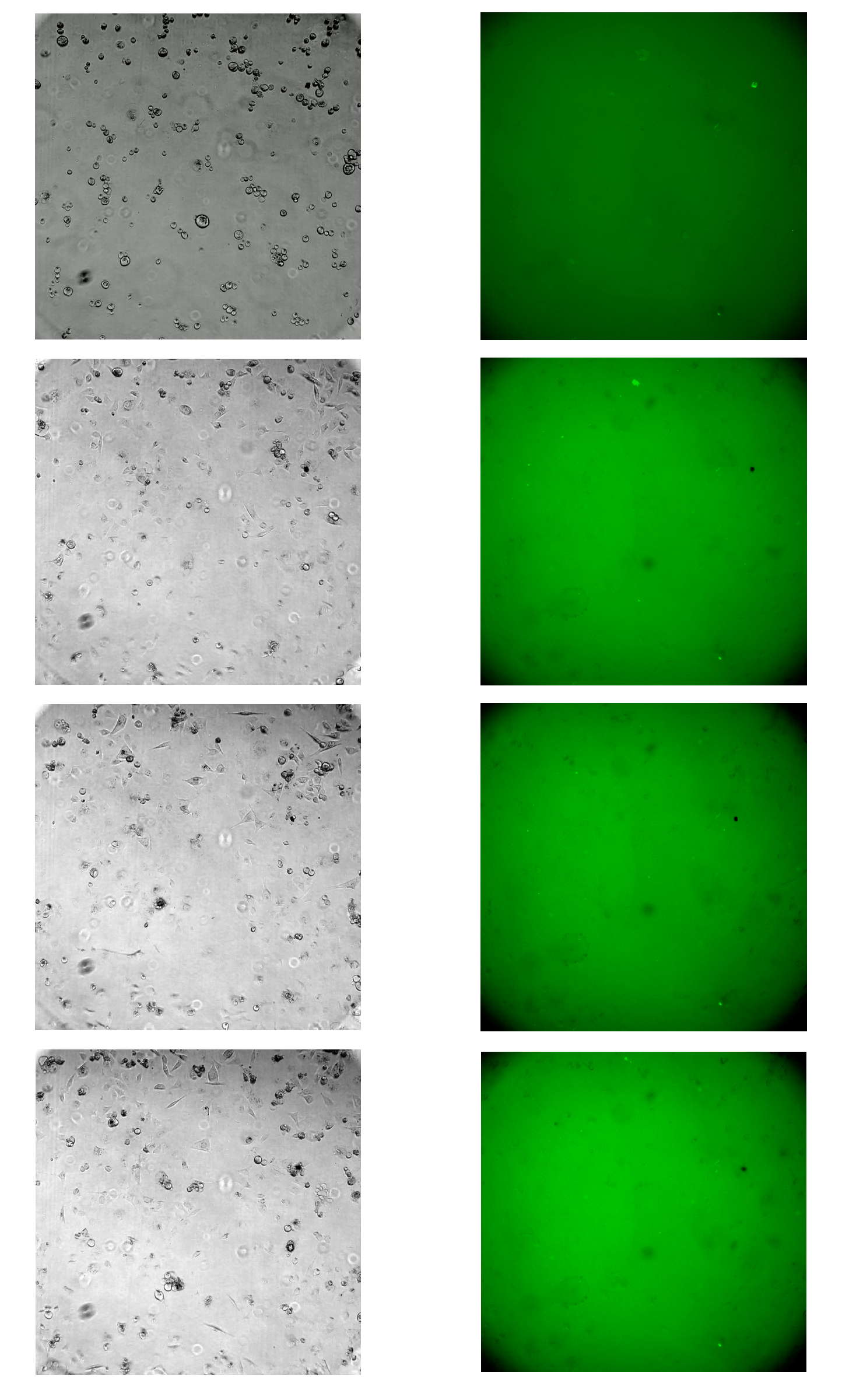}
				\caption{{Example of experimental data. On the left, cells position from the initial time to $72$ hours, pictured {every }day, on the right the corresponding gelatin consumption. Gelatin is in green, whereas {darker regions} indicate its degradation.}}
				\label{fig:73}
			\end{center}
		\end{figure}
		
		\begin{figure}[H]
			\centering
			\includegraphics[width=.75\textwidth]{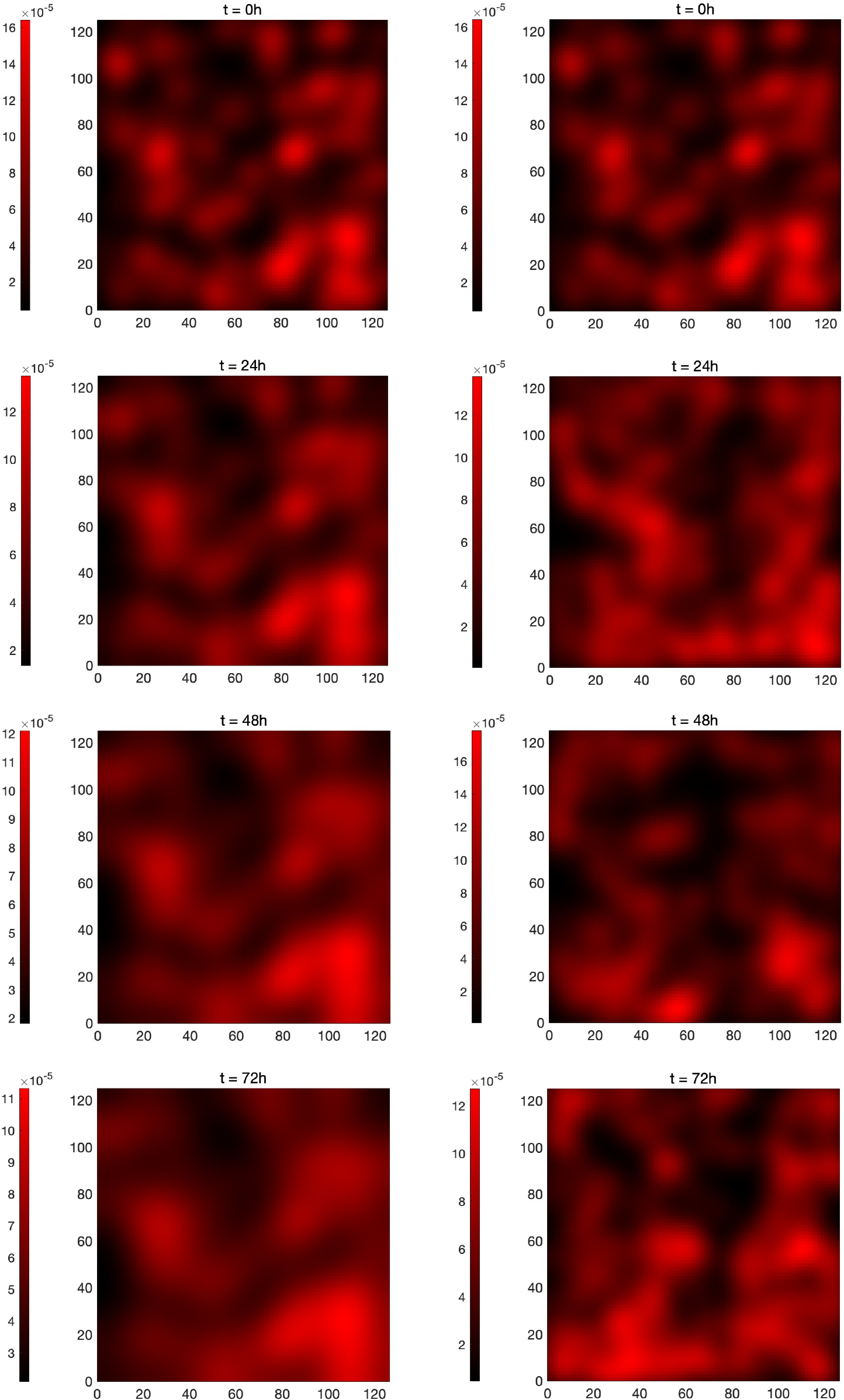}
			\caption{
				{Comparison between our numerical cell density with optimal parameter $\theta$ (on the left) and the experimental data after transformation into densities (on the right). 
					The domain dimension is in the corresponding {dimensionless} coordinates of the experimental one.
					In the first row, the initial data that is the same for both. The second row represents the density after one day, the third one after two days, whereas the last one after three days. We can observe some similarities between the two columns even if the match is not complete, but the error is below $0.4$. }}
			\label{fig:solnumcell}
		\end{figure}
		
		\begin{figure}[H]
			\centering
			\includegraphics[width=.75\textwidth]{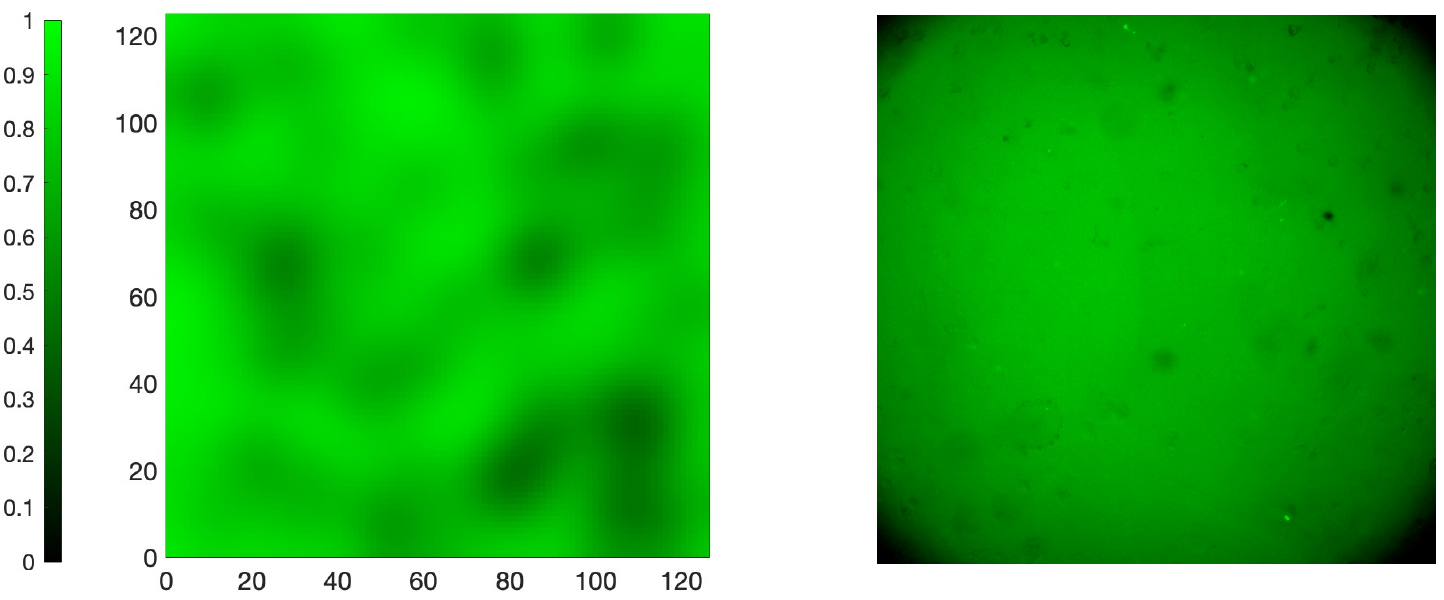}
			\caption{ {Gelatin degradation after $72$ hours. On the left the results obtained through our simulations, and on the right the experimental data. }}
			\label{fig:solnumgel72}
		\end{figure}

		
		\bibliographystyle{abbrvnat}
		\bibliography{references}	

\begin{thebibliography}{36}
\providecommand{\natexlab}[1]{#1}
\providecommand{\url}[1]{\texttt{#1}}
\expandafter\ifx\csname urlstyle\endcsname\relax
  \providecommand{\doi}[1]{doi: #1}\else
  \providecommand{\doi}{doi: \begingroup \urlstyle{rm}\Url}\fi

\bibitem[Aceto et~al.(2014)Aceto, Bardia, Miyamoto, Donaldson, Wittner,
  Spencer, Yu, Pely, Engstrom, Zhu, et~al.]{aceto}
N.~Aceto, A.~Bardia, D.~T. Miyamoto, M.~C. Donaldson, B.~S. Wittner, J.~A.
  Spencer, M.~Yu, A.~Pely, A.~Engstrom, H.~Zhu, et~al.
\newblock Circulating tumor cell clusters are oligoclonal precursors of breast
  cancer metastasis.
\newblock \emph{Cell}, 158\penalty0 (5):\penalty0 1110--1122, 2014.
\newblock URL \url{https://doi.org/10.1016/j.cell.2014.07.013}.

\bibitem[Aregba-Driollet et~al.(2004)Aregba-Driollet, Diele, and
  Natalini]{aregba}
D.~Aregba-Driollet, F.~Diele, and R.~Natalini.
\newblock A mathematical model for the sulphur dioxide aggression to calcium
  carbonate stones: Numerical approximation and asymptotic analysis.
\newblock \emph{SIAM Journal on Applied Mathematics}, 64\penalty0 (5):\penalty0
  1636--1667, 2004.
\newblock URL \url{https://doi.org/10.1137/S003613990342829X}.

\bibitem[Artym et~al.(2006)Artym, Zhang, Seillier-Moiseiwitsch, Yamada, and
  Mueller]{artym}
V.~V. Artym, Y.~Zhang, F.~Seillier-Moiseiwitsch, K.~M. Yamada, and S.~C.
  Mueller.
\newblock Dynamic interactions of cortactin and membrane type 1 matrix
  metalloproteinase at invadopodia: defining the stages of invadopodia
  formation and function.
\newblock \emph{Cancer Res.}, 66\penalty0 (6):\penalty0 3034--3043, 2006.
\newblock URL \url{https://doi.org/10.1158/0008-5472.CAN-05-2177}.

\bibitem[Braun(2021)]{braunphd}
E.~C. Braun.
\newblock \emph{Organs-On-Chips: mathematical modelling and parameter
  estimation}.
\newblock PhD thesis, Università degli Studi di Roma Tre, 2021.
\newblock URL
  \url{http://www.matfis.uniroma3.it/Allegati/Dottorato/TESI/ecbraun/Thesis%20Braun%20Revision.pdf}.

\bibitem[Braun et~al.(2022)Braun, Bretti, and Natalini]{braunarticle}
E.~C. Braun, G.~Bretti, and R.~Natalini.
\newblock Parameter estimation techniques for a chemotaxis model inspired by
  cancer-on-chip (coc) experiments.
\newblock \emph{Int. J. Non Linear Mech.}, 140:\penalty0 103895, 2022.
\newblock URL \url{https://doi.org/10.1016/j.ijnonlinmec.2021.103895}.

\bibitem[Bubba et~al.(2019)Bubba, Pouchol, Ferrand, Vidal, Almeida, Perthame,
  and Sabbah]{bubba}
F.~Bubba, C.~Pouchol, N.~Ferrand, G.~Vidal, L.~Almeida, B.~Perthame, and
  M.~Sabbah.
\newblock A chemotaxis-based explanation of spheroid formation in 3d cultures
  of breast cancer cells.
\newblock \emph{J. Theor. Biol.}, 479:\penalty0 73--80, 2019.
\newblock URL \url{https://doi.org/10.1016/j.jtbi.2019.07.002}.

\bibitem[Chaplain et~al.(2019)Chaplain, Giverso, Lorenzi, and
  Preziosi]{giverso}
M.~A.~J. Chaplain, C.~Giverso, T.~Lorenzi, and L.~Preziosi.
\newblock Derivation and application of effective interface conditions for
  continuum mechanical models of cell invasion through thin membranes.
\newblock \emph{SIAM J. Appl. Math.}, 79\penalty0 (5):\penalty0 2011--2031,
  2019.
\newblock ISSN 0036-1399.
\newblock \doi{10.1137/19M124263X}.
\newblock URL \url{https://doi.org/10.1137/19M124263X}.

\bibitem[Chen and Singer(1980)]{chen1}
W.-T. Chen and S.~Singer.
\newblock Fibronectin is not present in the focal adhesions formed between
  normal cultured fibroblasts and their substrata.
\newblock \emph{Proc. Natl. Acad. Sci.}, 77\penalty0 (12):\penalty0 7318--7322,
  1980.
\newblock URL \url{https://doi.org/10.1073/pnas.77.12.7318}.

\bibitem[Chen et~al.(1984)Chen, Olden, Bernard, and Chu]{chen2}
W.-T. Chen, K.~Olden, B.~A. Bernard, and F.~F. Chu.
\newblock Expression of transformation-associated protease (s) that degrade
  fibronectin at cell contact sites.
\newblock \emph{J. Cell. Biol.}, 98\penalty0 (4):\penalty0 1546--1555, 1984.
\newblock URL \url{https://doi.org/10.1083/jcb.98.4.1546}.

\bibitem[Chen et~al.(1985)Chen, Chen, Parsons, and Parsons]{chen3}
W.-T. Chen, J.-M. Chen, S.~J. Parsons, and J.~T. Parsons.
\newblock Local degradation of fibronectin at sites of expression of the
  transforming gene product pp60src.
\newblock \emph{Nature}, 316\penalty0 (6024):\penalty0 156--158, 1985.
\newblock URL \url{https://doi.org/10.1038/316156a0}.

\bibitem[Ciavolella et~al.(2023)Ciavolella, David, and Poulain]{ciadapou}
G.~Ciavolella, N.~David, and A.~Poulain.
\newblock Effective interface conditions for a porous medium type problem.
\newblock \emph{to appear in Interface Free Bound}, 2023.
\newblock URL \url{https://arxiv.org/abs/2105.02063}.

\bibitem[Connolly and Maxwell(2002)]{connolly}
L.~Connolly and P.~Maxwell.
\newblock Image analysis of transwell assays in the assessment of invasion by
  malignant cell lines.
\newblock \emph{Br. J. Biomed. Sci.}, 59\penalty0 (1):\penalty0 11--14, 2002.
\newblock URL \url{https://doi.org/10.1080/09674845.2002.11783627}.

\bibitem[Di~Costanzo et~al.(2016)Di~Costanzo, Ingangi, Angelini, Carfora,
  Carriero, and Natalini]{dicostanzo}
E.~Di~Costanzo, V.~Ingangi, C.~Angelini, M.~F. Carfora, M.~V. Carriero, and
  R.~Natalini.
\newblock A macroscopic mathematical model for cell migration assays using a
  real-time cell analysis.
\newblock \emph{PLoS One}, 11\penalty0 (9):\penalty0 e0162553, 2016.
\newblock URL \url{https://doi.org/10.1371/journal.pone.0162553}.

\bibitem[Dillek{\aa}s et~al.(2019)Dillek{\aa}s, Rogers, and Straume]{dillekas}
H.~Dillek{\aa}s, M.~S. Rogers, and O.~Straume.
\newblock Are 90\% of deaths from cancer caused by metastases?
\newblock \emph{Cancer med.}, 8\penalty0 (12):\penalty0 5574--5576, 2019.
\newblock URL \url{https://doi.org/10.1002/cam4.2474}.

\bibitem[Franssen et~al.(2019)Franssen, Lorenzi, Burgess, and
  Chaplain]{franssen}
L.~C. Franssen, T.~Lorenzi, A.~E. Burgess, and M.~A. Chaplain.
\newblock A mathematical framework for modelling the metastatic spread of
  cancer.
\newblock \emph{Bull. Math. Biol.}, 81\penalty0 (6):\penalty0 1965--2010, 2019.
\newblock URL \url{https://doi.org/10.1007/s11538-019-00597-x}.

\bibitem[Friedl and Wolf(2010)]{friedl}
P.~Friedl and K.~Wolf.
\newblock Plasticity of cell migration: a multiscale tuning model.
\newblock \emph{J. Cell Biol.}, 188\penalty0 (1):\penalty0 11--19, 2010.
\newblock URL \url{https://doi.org/10.1083/jcb.200909003}.

\bibitem[Gallinato et~al.(2017)Gallinato, Colin, Saut, and Poignard]{gallinato}
O.~Gallinato, T.~Colin, O.~Saut, and C.~Poignard.
\newblock Tumor growth model of ductal carcinoma: from in situ phase to stroma
  invasion.
\newblock \emph{J. Theoret. Biol.}, 429:\penalty0 253--266, 2017.
\newblock URL \url{https://doi.org// 10.1016/j.jtbi.2017.06.022}.

\bibitem[Giverso et~al.(2022)Giverso, Lorenzi, and Preziosi]{giverso2}
C.~Giverso, T.~Lorenzi, and L.~Preziosi.
\newblock Effective interface conditions for continuum mechanical models
  describing the invasion of multiple cell populations through thin membranes.
\newblock \emph{Appl. Math. Lett.}, 125:\penalty0 107708, 2022.
\newblock URL \url{https://doi.org/10.1016/j.aml.2021.107708}.

\bibitem[Guarguaglini and Natalini(2007)]{guarguaglini}
F.~Guarguaglini and R.~Natalini.
\newblock Fast reaction limit and large time behavior of solutions to a
  nonlinear model of sulphation phenomena.
\newblock \emph{Communications in Partial Differential Equations}, 32\penalty0
  (2):\penalty0 163--189, 2007.
\newblock URL \url{https://doi.org/10.1080/03605300500361438}.

\bibitem[Harbeck et~al.(2019)Harbeck, Penault-Llorca, Cortes, Gnant, Houssami,
  Poortmans, Ruddy, Tsang, and Cardoso]{harbeck}
N.~Harbeck, F.~Penault-Llorca, J.~Cortes, M.~Gnant, N.~Houssami, P.~Poortmans,
  K.~Ruddy, J.~Tsang, and F.~Cardoso.
\newblock Breast cancer.
\newblock \emph{Nat. Rev. Dis. Primers}, 5\penalty0 (1):\penalty0 1--31, 2019.
\newblock URL \url{https://doi.org/10.1038/s41572-019-0111-2}.

\bibitem[Hong et~al.(2016)Hong, Fang, and Zhang]{hong}
Y.~Hong, F.~Fang, and Q.~Zhang.
\newblock Circulating tumor cell clusters: What we know and what we expect.
\newblock \emph{Int. J. Oncol.}, 49\penalty0 (6):\penalty0 2206--2216, 2016.
\newblock URL \url{https://doi.org/10.3892/ijo.2016.3747}.

\bibitem[Kachanov(1986)]{kachanov}
L.~Kachanov.
\newblock \emph{Introduction to continuum damage mechanics}, volume~10.
\newblock Springer Science \& Business Media, 1986.
\newblock URL \url{https://doi.org/10.1007/978-94-017-1957-5}.

\bibitem[Ke et~al.(2011)Ke, Wang, Xu, and Abassi]{ke}
N.~Ke, X.~Wang, X.~Xu, and Y.~A. Abassi.
\newblock The xcelligence system for real-time and label-free monitoring of
  cell viability.
\newblock In \emph{Mammalian cell viability}, pages 33--43. Springer, 2011.
\newblock URL \url{https://doi.org/10.1007/978-1-61779-108-6_6}.

\bibitem[Malaquin et~al.(2013)Malaquin, Vercamer, Bouali, Martien, Deruy,
  Wernert, Chwastyniak, Pinet, Abbadie, and Pourtier]{malaquin}
N.~Malaquin, C.~Vercamer, F.~Bouali, S.~Martien, E.~Deruy, N.~Wernert,
  M.~Chwastyniak, F.~Pinet, C.~Abbadie, and A.~Pourtier.
\newblock Senescent fibroblasts enhance early skin carcinogenic events via a
  paracrine mmp-par-1 axis.
\newblock \emph{PloS One}, 8\penalty0 (5):\penalty0 e63607, 2013.
\newblock URL \url{https://doi.org/10.1371/journal.pone.0063607}.

\bibitem[Martin et~al.(2012)Martin, Hayes, Walk, Ammer, Markwell, and
  Weed]{martin}
K.~H. Martin, K.~E. Hayes, E.~L. Walk, A.~G. Ammer, S.~M. Markwell, and S.~A.
  Weed.
\newblock Quantitative measurement of invadopodia-mediated extracellular matrix
  proteolysis in single and multicellular contexts.
\newblock \emph{J. Vis. Exp.}, \penalty0 (66):\penalty0 e4119, 2012.
\newblock URL \url{https://dx.doi.org/10.3791/4119}.

\bibitem[Martinez-Serra et~al.(2014)Martinez-Serra, Gutierrez,
  Mu{\~n}oz-Cap{\'o}, Navarro-Palou, Ros, Amat, Lopez, Marcus, Fueyo, Suquia,
  et~al.]{martinez}
J.~Martinez-Serra, A.~Gutierrez, S.~Mu{\~n}oz-Cap{\'o}, M.~Navarro-Palou,
  T.~Ros, J.~C. Amat, B.~Lopez, T.~F. Marcus, L.~Fueyo, A.~G. Suquia, et~al.
\newblock xcelligence system for real-time label-free monitoring of growth and
  viability of cell lines from hematological malignancies.
\newblock \emph{OncoTargets Ther.}, 7:\penalty0 985—994, 2014.
\newblock URL \url{https://doi.org/10.2147/OTT.S62887}.

\bibitem[Morton and Mayers(2005)]{morton}
K.~W. Morton and D.~F. Mayers.
\newblock \emph{Numerical solution of partial differential equations: an
  introduction}.
\newblock Cambridge university press, 2005.
\newblock URL \url{https://doi.org/10.1017/CBO9780511812248}.

\bibitem[Murray(2002)]{murray1}
J.~D. Murray.
\newblock \emph{Mathematical Biology I. An Introduction}.
\newblock Springer, 2002.
\newblock URL \url{https://doi.org/10.1007/b98868}.

\bibitem[Obr et~al.(2013)Obr, R{\"o}selov{\'a}, Grebe{\v{n}}ov{\'a}, and
  Ku{\v{z}}elov{\'a}]{obr}
A.~Obr, P.~R{\"o}selov{\'a}, D.~Grebe{\v{n}}ov{\'a}, and K.~Ku{\v{z}}elov{\'a}.
\newblock Real-time monitoring of hematopoietic cell interaction with
  fibronectin fragment: The effect of histone deacetylase inhibitors.
\newblock \emph{Cell Adh. Migr.}, 7\penalty0 (3):\penalty0 275--282, 2013.
\newblock URL \url{https://doi.org/10.4161/cam.24531}.

\bibitem[Quarteroni et~al.(2010)Quarteroni, Sacco, and Saleri]{qss}
A.~Quarteroni, R.~Sacco, and F.~Saleri.
\newblock \emph{Numerical mathematics}.
\newblock Springer Science \& Business Media, 2010.
\newblock URL \url{https://doi.org/10.1007/b98885}.

\bibitem[Saltelli et~al.(2008)Saltelli, Ratto, Andres, Campolongo, Cariboni,
  Gatelli, Saisana, and Tarantola]{saltelli}
A.~Saltelli, M.~Ratto, T.~Andres, F.~Campolongo, J.~Cariboni, D.~Gatelli,
  M.~Saisana, and S.~Tarantola.
\newblock \emph{Global sensitivity analysis. The Primer}.
\newblock John Wiley \& Sons, 2008.
\newblock URL \url{https://doi.org/10.1002/9780470725184}.

\bibitem[Segel(1972)]{segel}
L.~A. Segel.
\newblock Simplification and scaling.
\newblock \emph{SIAM review}, 14\penalty0 (4):\penalty0 547--571, 1972.
\newblock URL \url{https://doi.org/10.1137/1014099}.

\bibitem[Sung et~al.(2021)Sung, Ferlay, Siegel, Laversanne, Soerjomataram,
  Jemal, and Bray]{sung}
H.~Sung, J.~Ferlay, R.~L. Siegel, M.~Laversanne, I.~Soerjomataram, A.~Jemal,
  and F.~Bray.
\newblock Global cancer statistics 2020: Globocan estimates of incidence and
  mortality worldwide for 36 cancers in 185 countries.
\newblock \emph{CA Cancer J. Clin.}, 71\penalty0 (3):\penalty0 209--249, 2021.
\newblock URL \url{https://doi.org/10.3322/caac.21660}.

\bibitem[T{\"u}rker~{\c{S}}ener et~al.(2017)T{\"u}rker~{\c{S}}ener, Albeniz,
  Din{\c{c}}, and Albeniz]{turker}
L.~T{\"u}rker~{\c{S}}ener, G.~Albeniz, B.~Din{\c{c}}, and I.~Albeniz.
\newblock icelligence real-time cell analysis system for examining the
  cytotoxicity of drugs to cancer cell lines.
\newblock \emph{Exp. Ther. Med.}, 14\penalty0 (3):\penalty0 1866--1870, 2017.
\newblock URL \url{https://doi.org/10.3892/etm.2017.4781}.

\bibitem[Waks and Winer(2019)]{waks}
A.~G. Waks and E.~P. Winer.
\newblock Breast cancer treatment: a review.
\newblock \emph{Jama}, 321\penalty0 (3):\penalty0 288--300, 2019.
\newblock URL \url{https://doi.org/10.1001/jama.2018.19323}.

\bibitem[Zaoui et~al.(2019)Zaoui, Morel, Ferrand, Fellahi, Bastard,
  Lamazi{\`e}re, Larsen, B{\'e}r{\'e}ziat, Atlan, and Sabbah]{zaoui}
M.~Zaoui, M.~Morel, N.~Ferrand, S.~Fellahi, J.-P. Bastard, A.~Lamazi{\`e}re,
  A.~K. Larsen, V.~B{\'e}r{\'e}ziat, M.~Atlan, and M.~Sabbah.
\newblock Breast-associated adipocytes secretome induce fatty acid uptake and
  invasiveness in breast cancer cells via cd36 independently of body mass
  index, menopausal status and mammary density.
\newblock \emph{Cancers}, 11\penalty0 (12):\penalty0 2012, 2019.
\newblock URL \url{https://doi.org/10.3390/cancers11122012}.

\end{thebibliography}
		
		\pagebreak
		\begin{center}
			\textbf{\large Supplementary Materials}
		\end{center}
		\setcounter{equation}{0}
		\setcounter{figure}{0}
		\setcounter{table}{0}
		\setcounter{page}{1}
		\makeatletter
		\renewcommand{\thesection}{S\arabic{section}}
		\renewcommand{\theequation}{S\arabic{equation}}
		\renewcommand{\thefigure}{S\arabic{figure}}
		\renewcommand{\bibnumfmt}[1]{[S#1]}
		\renewcommand{\citenumfont}[1]{S#1}
		
		
		\section{Discretisation}\label{app:discretisation}
		We illustrate the two-dimension numerical method, Morton and Mayers~\cite{morton}, Quarteroni \textit{et al.}~\cite{qss}.  
		We present the discretisation of the two-dimensional System~\eqref{eq:nondim} on the domain $\Omega=[a,b]\times [c,d]$. 
		
		Concerning space discretisation, we consider a mesh over $\Omega$ such that $\Delta x = \frac{b-a}{N_x+1}= \Delta y$. Thus, we divide $[a,b]$ into $N_x \in \N$ intervals such that $a$ (respectively, $b$) corresponds to $j=1$ ($j=N_x+1$), and $[c,d]$ into $N_y \in \N$ intervals such that $c$ (respectively, $d$) corresponds to $i=1$ ($i=N_y+1$). 
		The mesh is formed by the intervals
		$$J_j= \left(x_{j-\frac 1 2}, x_{j+\frac 1 2}\right), \; j=1,...,N_x+1, \qquad I_i= \left(y_{i-\frac 1 2}, y_{i+\frac 1 2}\right), \; i=1,...,N_y+1.$$
		The intervals are centred in $x_j= j \Delta x$, $j=1,...,N_x+1$ and $y_i= i \Delta y$, $i=1,...,N_y+1$. Moreover, we add ghost points to build the extremal intervals centred at the boundaries for $x=a,b$ and $y=c,d$.
		At a given time, the spatial discretisation of $u(t,x,y)$ (the same for functions $m$ and $d$), interpreted in the finite volume sense, is of the form
		\[u_{i,j}(t) \approx \frac{1}{\Delta x}\frac{1}{\Delta y} \int_{J_j}\int_{I_i} u(t,x,y) \, dx \, dy , \quad \mbox{ for } j=1,...,N_x+1, \, i=1,...,N_y+1.\] 
		
		For the time discretisation, we consider a time step $\Delta t$, and set $t^n = n \Delta t$, with $n\in \N$. The discrete approximation of $u(t,x,y)$ (or of functions $m$ or $d$), for $n\in \N$, $j=1,...,N_x+1$, $i=1,...,N_y+1$ is now 
		\[u^n_{i,j} \approx \frac{1}{\Delta x}\frac{1}{\Delta y} \int_{J_j}\int_{I_i} u(t^n,x,y) \, dx \, dy.
		\]
		
		\textbf{Equation for $\mathbf{d}$.} We solve $\pt d =\frac{1}{p} m \left(1-\frac 1 p d\right)$. We discretise the solution $d=\frac 1 p \left(1-e^{\int_0^t m ds}\right),$ $\forall x,y\in\Omega$, using the trapezoidal rule for the integral 
		\[\int_0^t m \, ds\approx \frac{\Delta t}{2} \left[m^n_{i,j}+2\sum_{k=1}^{n-1} m^k_{i,j}\right],\]
		for $j=1,...,N_x+1$, $i=1,...,N_y+1$.
		We remember that initially $d^0_{i,j}=0, \, \forall i=1,...,N_y+1, j=1,...,N_x+1$, since the gelatin is not damaged. \\[1ex]
		
		\textbf{Discretisation for $\mathbf{D(d)}$.} We analyse separately the diffusion coefficient for cell density $u$. We recall the expression in~\eqref{eq:diff}, namely $D(d)=\theta + d$ in the parametrised version. Its discretisation is easy, but we have to deal with ghost points that appear in the discretisation for $u$. We derived boundary conditions which preserve the mass in our system. Namely, we obtain
		\begin{equation}\label{eq:ghostD}
			\begin{array}{l}
				D(d)^n_{i,0}= D(d)^n_{i,2}, \quad D(d)^n_{i,N_x+2}= D(d)^n_{i,N_x},\\[1.5ex]
				D(d)^n_{0,j}= D(d)^n_{2,j}, \quad D(d)^n_{N_y+2,j}= D(d)^n_{N_y,j}.
			\end{array}
		\end{equation}
		
		\textbf{Equation for $\mathbf{u}$.}
		To achieve the time discretisation, we adopt an Euler method, and we write $\frac{du}{dt}(t)$ as $\frac{u^{n+1}-u^n}{\Delta t}$.  To obtain space discretisation, we recall the following explicit one-dimensional method
		\[
		\px(a(x) \px u) \approx \frac{(a_{j+1}+a_{j})(u_{j+1}-u_j) - (a_{j-1}+a_{j})(u_j-u_{j-1})}{2\Delta x^2}.
		\]
		In the vector analysis, we can use this one-dimensional approximation on each axis derivative.
		Then, full discretisation of System~\eqref{eq:nondim} reads for $n\in\N, j=1,...,N_x+1, i=1,...,N_y+1$ as
		\begin{equation}
			\begin{array}{c} u^{n+1}_{i,j}-u^n_{i,j}=\mu[(D^n_{i-1,j}+D^n_{i,j})u^n_{i-1,j}+ (D^n_{i+1,j}+D^n_{i,j}) u^n_{i+1,j}+ \\[1ex]
				-(D^n_{i-1,j}+D^n_{i+1,j}+4 D^n_{i,j}+D^n_{i,j-1}+D^n_{i,j+1}) u^n_{i,j} + (D^n_{i,j-1}+D^n_{i,j}) u^n_{i,j-1} +  \\[1ex]
				+ (D^n_{i,j+1}+D^n_{i,j}) u^n_{i,j+1}]
				+k_2 u^n_{i,j} \left(1-\frac{u^n_{i,j}}{\alpha_3}\right),
			\end{array}
		\end{equation} 
		with $\mu_u =\frac{\Delta t}{2 \Delta x ^2}$.
		Finally, we deduce the system
		\begin{equation}\label{eq:discr_u}
			\begin{array}{ll}
				u^{n+1}_{i,j}&= [1-\mu (D^n_{i-1,j}+D^n_{i+1,j}+4 D^n_{i,j}+D^n_{i,j-1}+D^n_{i,j+1})] u^n_{i,j}+ \\[1ex] 
				&+ \mu(D^n_{i-1,j}+D^n_{i,j})u^n_{i-1,j}+\mu (D^n_{i+1,j}+D^n_{i,j}) u^n_{i+1,j}+\mu (D^n_{i,j-1}+D^n_{i,j}) u^n_{i,j-1} +\\[1ex]
				&+\mu (D^n_{i,j+1}+D^n_{i,j}) u^n_{i,j+1}+k_2 u^n_{i,j} \left(1-\frac{u^n_{i,j}}{\alpha_3}\right),
			\end{array}
		\end{equation}
		with second order discretisation of the Neumann boundary conditions in $x=a, b$, and $y= c,d$ as
		\begin{equation}\label{eq:discr_N}
			u^{n}_{i,0}=u^{n}_{i,2}, \quad u^{n}_{i,N_x+2}= u^{n}_{i,N_x}, \quad u^{n}_{0,j}=u^{n}_{2,j}, \quad u^{n}_{N_y+2,j}= u^{n}_{N_y,j}.
		\end{equation}
		Previous conditions give the relations of the extremal ghost points. 
		Substituting ghost points Relations~\eqref{eq:ghostD},~\eqref{eq:discr_N} into Equation~\eqref{eq:discr_u}, we obtain the discretised equation for $u$ of System~\eqref{eq:nondim} defined on the spatial grid. \\[1ex]
		
		\textbf{Equation for $\mathbf{m}$.}
		Instead of the equation for $m$, in order to avoid stiffness problems due to the presence of the term $-m$, we discretise the equation for $w=e^{t}m$ which is of the form $\pt w= \Delta w + k_1 u (1-pd) e^{ t}$. At the end, the discretised density for the enzymes $m$ can be derived from the numerical solution $w$, as $m^n_{i,j}=e^{-t^n} w^n_{i,j}$. We infer that
		\begin{equation}\label{eq:discr_m}
			w^{n+1}_{i,j}= (1-4\mu_m)w^n_{i,j}+\mu_m (w^n_{i-1,j}+w^n_{i+1,j}+w^n_{i,j-1}+w^n_{i,j+1})+k_1 u^n_{i,j}(1-p d^n_{i,j}) e^{ n \Delta t},
		\end{equation}
		with $\mu_m=\frac{\Delta t }{\Delta x^2}$. Boundary conditions are the same as in \eqref{eq:discr_N}.\\[1ex]
		
		Calling the vector solutions at time $t^n$ as
		$$U^n = \begin{array}{cccccc}
			\left(u^n_{i,1}, \ldots, u^n_{i,N_x+1}\right)^T,
		\end{array} \qquad
		W^n =\begin{array}{cccccc}
			\left(w^n_{i,1}, \ldots, w^n_{i,N_x+1}\right)^T,
		\end{array}$$
		and the reaction vectors as
		\[R^{n} = k_2 U^n \left(1-\frac{U^n}{\alpha_3}\right),\quad
		S^{n} = e^{\alpha n \Delta t} \beta U^n,\]
		we can write the discretised systems in a matrix form as $ U^{n+1}= A U^n + \Delta t R^n$ coupled with $W^{n+1}= B W^n + \Delta t S^n$. Coefficients of the matrices $A$ and $B$ can be found substituting ghost points in the discretised equations. In particular, $B$ is the standard matrix related to the heat equation. 
		
		It is important to guarantee the positiveness of the coefficients in the previous Equations~\eqref{eq:discr_u}, \eqref{eq:discr_m}, in order to preserve positiveness and stability. This brings conditions on the time interval $\Delta t$ which has to be such that
		\begin{equation}
			\Delta t < \min(dt_1,dt_2), \quad \mbox{ where } \quad dt_1= \frac{\Delta x^2}{4 (\theta +1)}, \quad dt_2= \frac{\Delta x^2}{4},
		\end{equation}
		since $dt_1= \min\left(\frac{2 \Delta x^2}{8\theta+(d^n_{i-1,j}+d^n_{i+1,j}+4 d^n_{i,j}+d^n_{i,j-1}+d^n_{i,j+1})}\right)= \left(\frac{2 \Delta x^2}{8\theta+ \max(d^n_{i-1,j}+d^n_{i+1,j}+4 d^n_{i,j}+d^n_{i,j-1}+d^n_{i,j+1})}\right)$, and the maximum value for $d$ is $1$.

	\end{document}